\documentstyle[eqsecnum,aps,prd]{revtex}
\def\be{\begin{equation}}
\def\ee{\end{equation}}
\def\ba{\begin{eqnarray}}
\def\ea{\end{eqnarray}}

\def\tbox{\stackrel{\sim}{\Box}}
\def\teps{{\widetilde{\epsilon}}^{\mu\nu\lambda\kappa}}
\def\tnabla{\widetilde{\nabla}}
\def\trho{\widetilde{\rho}}
\def\ta{\widetilde{a}}

\def\tg{\widetilde{g}}
\def\tH{\widetilde{H}}

\def\tR{\widetilde{R}}

\def\tsigma{\widetilde{\sigma}}
\def\tt{\widetilde{t}}
\def\tT{\widetilde{T}}
\def\ttheta{\widetilde{\theta}}
\def\tts{\left(\displaystyle{t\over{\ t_*}}\right)}
\def\tTbeta{\ ^{(\beta)}\widetilde{T}}
\def\tTH{\ ^{(H)}\widetilde{T}}
\def\tTphi{\ ^{(\phi)}\widetilde{T}}

\def\del{\partial}
\def\tal{{\widetilde\alpha}}
\def\lone{\lambda - \lambda_1}
\def\ltwo{\lambda - \lambda_2}
\def\lthree{\lambda - \lambda_3}
\def\lzero{\lambda - \lambda_0}
\def\lzerop{\lambda - \lambda'_0}
\def\bzero{{\widehat{B}_{12}}}
\def\sech{{\rm sech}}

\def\half{\frac{1}{2}}

\def\l{\label}

\begin{document}
  \begin{flushright}
SUSSEX-AST-94/10-1\\
hep-th/9410136\\
  \end{flushright}
  \begin{center}
\vspace{1 in}
\Large
{\bf String Cosmology with a Time-Dependent \\
Antisymmetric Tensor Potential\\}
\vspace{0.2 in}
\normalsize
\large{E. J.~Copeland, Amitabha Lahiri \& David Wands} \\
\normalsize
\vspace{0.2 in}
{\em  School of Mathematical \& Physical Sciences, \\
University of Sussex, \\ Brighton BN1 9QH.\\U.~K.}\\
\end{center}
\vspace{0.2 in}

\begin{abstract}

We present a class of exact solutions for homogeneous, anisotropic
cosmologies in four dimensions derived from the low-energy string
effective action including a homogeneous dilaton $\phi$ and
antisymmetric tensor potential $B_{\mu\nu}$.  Making this potential
time-dependent produces an anisotropic energy-momentum tensor, and
leads us to consider a Bianchi I cosmology. The solution for the axion
field must then only be a linear function of one spatial coordinate.
This in turn places an upper bound on the product of the two scale
factors evolving perpendicular to the gradient of the axion field.
The only late-time isotropic solution is then a {\em contracting}
universe.

\end{abstract}
\vspace{0.2 in}


\section{Introduction}
\l{sectINTRO}

The massless excitations of a string consist not only of the graviton
field, $g_{\mu\nu}$, of general relativity, but also a dilaton field,
$\phi$, which determines the strength of the gravitational coupling,
and an antisymmetric tensor potential, $B_{\mu\nu}$. While the
cosmological consequences of the dilaton have been extensively
discussed \cite{Gas+Ven 93,dilaton}, the role of the antisymmetric
tensor field strength
\be
H_{\mu\nu\lambda}=\partial_{[\mu}B_{\nu\lambda]}
\ee
is often less clear.  This is partly due to the difficulty of handling
the many new degrees of freedom this introduces in higher dimensions.
Here we will consider the field restricted to a four-dimensional
cosmology where we have only one degree of freedom which can be
represented by a pseudo-scalar ``axion'' field.  The omission of
$H_{\mu\nu\lambda}$ is often justified due to the existence of duality
transforms of the string action which relate the dilaton-only
solutions to non-trivial $H$ field solutions, but the complete
equivalence of the solutions is only true if this duality extends to
the full action. In a previous paper \cite{CLW/6/94} we gave exact
solutions of the lowest order string $\beta$ function equations for
four-dimensional cosmologies with a time-dependent axion field (see
also \cite{axion}) which are related to the homogeneous dilaton-vacuum
cosmologies by an $SL(2,R)$ transform \cite{SL2R}.

Another commonly invoked symmetry is the $O(d,d)$ duality
\cite{Mei+Ven 91,Sen 91} which requires (in a cosmological setting)
both metric and antisymmetric potential to be functions only of time.
Here we will give explicit solutions for cosmologies including a
time-dependent $B_{\mu\nu}$, which can be seen to preclude any
time-dependence of $H_{\mu\nu\lambda}$ from its definition.  We shall
show that, as it is the field $H$ that appears in the metric equations
of motion, this is a highly restrictive prescription.  In particular
it introduces an anisotropic energy-momentum tensor which we shall
show inevitably leads to an anisotropic cosmology.

We will solve the string $\beta$ function equations only to lowest
order, which can be derived from the low-energy effective
action of the bosonic sector of a string theory reduced to four
dimensions \cite{effaction};
\be
S = \frac{1}{2\kappa^2} \int d^4x \sqrt{-g} e^{-\phi} \left[
	R + \left( \nabla\phi \right)^2 - n\left(\nabla\beta\right)^2
        - V
	- \frac{1}{12} H^2 \right]
\l{SACT}
\ee
where $H^2=H_{\mu\nu\lambda} H^{\mu\nu\lambda}$, $\kappa^2=8\pi G$ and
the modulus field $\beta$ represents the evolution of $n$ compact
dimensions. For simplicity we assume that these dimensions are
described by a spatially flat (Bianchi type I) metric with scale
factors $b_i$, and we define
$n\dot\beta^2\equiv\sum_i(\dot{b}_i/b_i)^2$.
We have adopted the sign conventions denoted (+++) by Misner, Thorne
and Wheeler \cite{MTW}.  The constant $V$ is proportional to the
central charge of the string theory.

The effect of certain types of ``stringy matter'' has been considered
elsewhere in the literature \cite{Gas+Ven 93}, where an equation of
state for the matter was assumed.
The symmetries of the vacuum, as well as any additional gauge
symmetries that may be present, will affect the matter Lagrangian as
well, and also the value of the central charge $V$. We shall assume
that the original string theory, from which the effective action
Eq.~(\ref{SACT}) is derived, contrives to set the central charge
$V = 0$, by adding appropriate bosonic or fermionic conformal matter.
Initially we shall restrict ourselves to vacuum solutions as regards
these matter fields in order to
examine the dynamical effect of the bosonic fields. Later we will
briefly discuss the possible effect of other matter, in particular
radiation.

The field equations are derived by varying this action (with $V=0$)
with respect to $g_{\mu\nu}$, $B_{\mu\nu}$ and $\phi$, respectively,
\ba
R_{\mu}^{\nu} - \frac{1}{2} g_{\mu}^{\nu} R & = &
	\frac{1}{12} \left( 3H_{\mu\lambda\kappa}H^{\nu\lambda\kappa}
			- \frac{1}{2} g_{\mu}^{\nu} H^2 \right)
	+ n \left( g_\mu^\lambda g^{\nu\kappa}
		- \half g_\mu^\nu g^{\lambda\kappa} \right)
			\nabla_\lambda\beta \nabla_\kappa\beta
	\nonumber \\
& & \quad - \frac{1}{2} g_{\mu}^{\nu} \left( \nabla\phi \right)^2
	+ \left( g_{\mu}^{\nu} g^{\lambda\kappa}
			-g_{\mu}^{\lambda} g^{\nu\kappa} \right)
		\nabla_\lambda\nabla_\kappa\phi
 \; , \l{SGMU} \\
\nabla_{\mu} \left( e^{-\phi} H^{\mu\nu\lambda} \right) & = & 0 \; ,
 \label{Heom} \l{SH} \\
\nabla_{\mu} \left( e^{-\phi} \nabla^\mu \beta \right) & = & 0 \; \\
2\Box\phi
 & = &  -R + \left(\nabla\phi\right)^2 + n\left(\nabla\beta\right)^2
	+ \frac{1}{12}H^2 \; .
 \l{SDIL}
\ea

These equations can be re-written in a more familiar general
relativistic form in terms of the conformally transformed Einstein
metric, defined as
\be
\tg_{\mu\nu} = e^{-\phi} g_{\mu\nu} \; .
\l{EMET}
\ee
In terms of this metric, the action (with $V=0$)
appears as the Einstein--Hilbert
action of general relativity while the dilaton appears simply as a
matter field, albeit one interacting with the other matter fields.
\be
S = \frac{1}{2\kappa^2} \int d^4x \sqrt{-\tg} \left[
	\tR - \half \left( \tnabla\phi \right)^2
	- n \left( \tnabla\beta \right)^2
	- \frac{1}{12} e^{-2\phi} \tH^2 \right] .
\l{EACT}
\ee
In this expression, raising of the indices was done with the inverse
$\tg^{\mu\nu}$ of the transformed metric (\ref{EMET}).
Note that $\tH_{\mu\lambda\kappa} \equiv H_{\mu\lambda\kappa}$, the
definition being metric-independent.

The corresponding field equations are then those for interacting
fields in general relativity;
\ba
\tR_{\mu\nu} - \frac{1}{2} \tg_{\mu\nu} \tR & = &
 \kappa^2 \left( \tTH_{\mu\nu} + \tTbeta_{\mu\nu}
 + \tTphi_{\mu\nu} \right)\; ,
 \l{EGMU} \label{einstein} \\
\tnabla_{\mu} \left( e^{-2\phi} \tH^{\mu\nu\lambda} \right)
 & = & 0 \; ,
 \l{EH} \l{einsteinHeom} \\
\tbox\beta & = & 0 \l{EBETA} \\
\tbox\phi + \frac{1}{6} e^{-2\phi} \tH^2 & = & 0 .
 \l{EDIL} \l{einsteinphieom}
\ea
The energy-momentum tensors appearing on the right-hand side
of the Einstein equations correspond to the energy-momentum tensors
for the dilaton, moduli and H-fields respectively,
\ba
\kappa^2 \tTphi_{\mu}^{\nu} & = & \half
	\left( \tg_{\mu}^{\lambda} \tg^{\nu\kappa} -
		\half \tg_{\mu}^{\nu} \tg^{\lambda\kappa} \right)
	\tnabla_\lambda\phi\tnabla_\kappa\phi \; , \l{ETDIL} \\
\kappa^2 \tTbeta_{\mu}^{\nu} & = & n
	\left( \tg_{\mu}^{\lambda} \tg^{\nu\kappa} -
		\half \tg_{\mu}^{\nu} \tg^{\lambda\kappa} \right)
	\tnabla_\lambda\beta\tnabla_\kappa\beta \; , \l{ETBETA} \\
\kappa^2 \tTH_{\mu}^{\nu} & = & \frac{1}{12}
e^{-2\phi} \left( 3\tH_{\mu\lambda\kappa}\tH^{\nu\lambda\kappa}
		-\frac{1}{2}\tg_{\mu}^{\nu}\tH^2 \right)
\; . \l{ETH}
\ea
While the total energy-momentum must be conserved (as guaranteed
by the Ricci identity) there are interactions between the three
components.

Because we assume that all fields are independent of the compact
dimensions we can immediately solve the equation of motion for the
antisymmetric tensor field, Eq.~(\ref{einsteinHeom}), in
four-dimensional spacetime by the Ansatz
\be
\tH^{\mu\nu\lambda} = e^{2\phi} \teps\tnabla_\kappa h,\l{ansatz}
\ee
where $\teps$ is the antisymmetric volume form in four dimensions
(obeying $\tnabla_{\rho}\teps=0$). The field $h$ obeys a new equation
of motion, derived from the integrability condition,
$\partial_{[\mu}\tH_{\nu\lambda\kappa]}=0$, which becomes
\be
\tbox\,h + 2 \tnabla^{\mu}\phi \tnabla_{\mu}h = 0 \; .
\label{einsteinheom}
\ee
We shall follow the usual string nomenclature and refer to $h$ as the
axion, even though its axion-like properties are not relevant to our
analysis. The effective energy-momentum tensor for the antisymmetric
tensor field in the Einstein frame can then be written as
\be
\kappa^2
 \tTH_{\mu}^{\nu} = \half
	\left( \tg_{\mu}^{\lambda} \tg^{\nu\kappa} -
		\half \tg_{\mu}^{\nu} \tg^{\lambda\kappa} \right)
	e^{2\phi} \tnabla_\lambda h \tnabla_\kappa h
\ee

Similarly the dilaton equation of motion, Eq.~(\ref{einsteinphieom}),
can be rewritten in terms of $h$ rather than of
$H_{\mu\nu\lambda}$,
\be
\tbox\,\phi = e^{2\phi} \left(\tnabla h\right)^2 \; .
\ee

\section{Solutions}

There are several possible homogeneous four-dimensional cosmologies
one may have in this system of dilaton and axion coupled to gravity.
The case where the axion is time-dependent was
discussed in a previous paper \cite{CLW/6/94}.  Here we consider
the case where the components of the antisymmetric tensor potential
$B_{\mu\nu}$ depend only on time, $B_{0i} = 0$ and $B_{ij} \equiv
B_{ij}(t)$.  Note that $B_{0i}$ can be always set to zero by utilising
the symmetry of the action under the vector gauge transformation
$B_{\mu\nu} \to\ B_{\mu\nu} + \del_{[\mu}\Lambda_{\nu]}$.
This is the case commonly discussed in the literature in the context of the
$O(d,d)$ symmetry of the low-energy action, Eq.~(\ref{SACT}), when the
metric and antisymmetric tensor potential are independent of $d=D-1$
of the spacetime coordinates.
As can be easily checked, for our Ansatz (\ref{ansatz}), a homogeneous
$B_{\mu\nu}$ corresponds to the situation where $\del_t h = 0$.

The modulus and dilaton fields are taken to be homogeneous as well;
$\beta\equiv\beta(t)$ and $\phi\equiv\phi(t)$. They then act like
stiff fluids in the Einstein frame with an isotropic pressure equal to
their density, so the energy-momentum tensors are
\be
\tT_\mu^\nu =
 {\rm diag}(-\trho,\trho,\trho,\trho) \ ,
\ee
where the energy densities are
\be
\trho_\beta = {n\over2\kappa^2}\dot\beta^2\; , \qquad
\trho_\phi = {1\over4\kappa^2}\dot\phi^2\; ,
\ee
and ``dot'' denotes $d/d\tt$.

We will consider a Bianchi I cosmology, $g_{\mu\nu} = {\rm diag}(-1,
a_1^2(t), a_2^2(t), a_3^2(t))$, the simplest form for an anisotropic
metric, where the homogeneous hypersurfaces of constant time have zero
spatial curvature.  We shall see that anisotropic expansion is
a necessary consequence of our choice of a homogeneous tensor
potential.

We solve first the evolution in the Einstein frame, (\ref{EMET}).
This metric is given by
\be
d\widetilde s^2 = - d\tt^2 + \ta_1^2(dx^1)^2 + \ta_2^2(dx^2)^2 +
\ta_3^2(dx^3)^2,
\ee
where we have defined $\ta_i = e^{-\phi/2}a_i$, and $d\tt =
e^{-\phi/2}dt$. Since our metric is diagonal, the equation of motion
for $h$, Eq.~(\ref{einsteinheom}), takes the form
\be
{1\over\ta_1^2}\del_1^2h + {1\over\ta_2^2}\del_2^2h +
{1\over\ta_3^2}\del_3^2h = 0.\l{hlaplace}
\ee
Also, from the off-diagonal components of Einstein equations,
Eq.~(\ref{einstein}), we have (since $\del_i\phi = 0$)
\be
0 = \half e^{2\phi}\del_ih\del_jh,\qquad\qquad(i, j = 1, 2,
3, i\neq j,{\rm no\ sum})
\l{LiLj}
\ee
In addition, by our assumption of homogeneity the stress-energy tensor
depends only on the time coordinate, which implies that the only
solution of Eq.~(\ref{hlaplace}) is
\be
\del_ih = L_i,
\ee
where $L_i, (i = 1,2,3),$ are constants, since $\del_th = 0$.  Then we
have $L_iL_j = 0,$ for all $i\neq j$, which tells us that only one of
$L_i$ can be non-zero. For definiteness, we choose $L_1 = L_2 = 0$.
The resulting energy-momentum tensor for the axion field is then
\be
\tTH_\mu^\nu =
 {\rm diag}\left(-\trho_H,-\trho_H,-\trho_H,\trho_H\right)
\ee
where
\be
\trho_H = \frac{e^{2\phi} L_3^2}{4\kappa^2\ta_3^2} \; .
\ee
The axion thus exerts an anisotropic pressure --- positive in the $x_3$
direction but negative pressure along $x_1$ and $x_2$.

The Einstein equations, Eqs.~(\ref{einstein}), then lead to the equations of
motion for the scale factors, $\tal_i = \ln\ta_i$,
\ba
\ddot\tal_1 + \dot\tal_1(\dot\tal_1 + \dot\tal_2 + \dot\tal_3) &=& 0,
\l{talone}\\
\ddot\tal_2 + \dot\tal_2(\dot\tal_1 + \dot\tal_2 + \dot\tal_3) &=& 0,
\l{taltwo}\\
\ddot\tal_3 + \dot\tal_3(\dot\tal_1 + \dot\tal_2 + \dot\tal_3) &=&
{e^{2\phi}L_3^2\over 2\ta_3^2},\l{talthree}
\ea
together with the constraint equation
\be
\dot\tal_1\dot\tal_2 + \dot\tal_2\dot\tal_3 + \dot\tal_3\dot\tal_1
 = {1\over4}\dot\phi^2 + {n\over2}\dot\beta^2
  + {1\over4}{e^{2\phi}L_3^2\over\ta_3^2},\l{ancon}
\ee
where dots denote differentiation with respect to time in the Einstein
frame, $\tt$.
The modulus and dilaton equations can be written as
\ba
\ddot\beta + \dot\beta(\dot\tal_1 + \dot\tal_2 + \dot\tal_3) & = & 0
 \ , \l{betaone} \\
\ddot\phi + \dot\phi(\dot\tal_1 + \dot\tal_2 + \dot\tal_3) & = &
 - {e^{2\phi}L_3^2\over\ta_3^2} \ . \l{dilone}
\ea
The axion field drives the evolution of $\phi$ and $\ta_3$ but leaves
$\ta_1$, $\ta_2$ and $\beta$ to evolve as ``free'' fields, subject
only to damping by the spatial expansion.

Let us introduce a new time coordinate $\lambda$ via the relation
\be
d\lambda \equiv {d\tt \over \ta_1\ta_2\ta_3}
	= {e^{\phi}dt \over a_1a_2a_3} .
\l{deflam}
\ee
In terms of this variable the above equations simplify considerably,
\ba
{d^2\over d\lambda^2}\tal_1 = {d^2\over d\lambda^2}\tal_2 &=&
{d^2\over d\lambda^2}\beta = {d^2\over d\lambda^2}(\tal_3
 + {1\over 2}\phi)
 = 0 \l{taeq} \ , \\
{d^2\phi\over d\lambda^2} &=& - e^{2\phi}L_3^2\ta_1^2\ta_2^2 \ .
 \l{diltwo}\ea
The equations for the scale factors and modulus can be readily solved,
\ba
\tal_1 &=\ C_1(\lone), \qquad\qquad a_1 &=\
\exp({1\over 2}\phi + C_1(\lone))
\l{a1sol}\\
\tal_2 &=\ C_2(\ltwo), \qquad\qquad a_2 &=\
\exp({1\over 2}\phi + C_2(\ltwo))
\l{a2sol}\\
\tal_3 + {1\over 2}\phi &=\ C_3(\lthree), \qquad\qquad a_3 &=\
\exp(C_3(\lthree)),
\l{a3sol}
\ea
and
\be
\beta = C_n (\lambda-\lambda_n) \ ,
\ee
where $C_i, \lambda_i (i = 1,2,3,n)$ are constants of integration.
The ``free'' fields $\tal_1$, $\tal_2$ and $\beta$ are monotonic
functions of time, while $\phi$ and $\tal_3$, both driven by the axion
field, are linked.  The conformal transform back to the string metric
cancels out this dependence of the third scale factor on the dilaton
leaving $a_3$ a ``free'' field, while it is the evolution of $a_1$ and
$a_2$ that becomes tied to the dilaton.

These expressions can be substituted into the constraint equation
(\ref{ancon}) to give
\be
\left({d\phi\over d\lambda}\right)^2 + 2\left({d\phi\over
d\lambda}\right)(C_1 + C_2) - 4C_1C_2 - 4C_3(C_1 + C_2) + 2nC_n^2 +
e^{2\phi}L_3^2\ta_1^2\ta_2^2 = 0.
\l{vsq}
\ee
Since the last term is necessarily non-negative, the requirement that
the dilaton $\phi$ be real translates to a constraint on the
constants $C_i$,
\be
C_0^2 \equiv 2(C_1+C_2+C_3)^2 - (C_1-C_2)^2 - 2C_3^2 -2nC_n^2
 \geq 0 \ . \l{czip}
\ee
We will choose $C_0$ to be non-negative. (It can only be zero when
$L_3=0$, corresponding to a vacuum solution.)

Note that from the definition of $\lambda$ in Eq.~(\ref{deflam}),
we have
\be
t - t_0 = {1\over C_1 + C_2 + C_3}
 \exp(C_1(\lone) + C_2(\ltwo) + C_3(\lthree)),\l{time}
\ee
where $t_0$ is a constant of integration which corresponds to an
arbitrarily chosen origin of proper time in the string frame.
Note that from Eq.~(\ref{czip}), it follows that $C_1 + C_2 + C_3
= 0$ only when $C_0=0$ and all the $C_i=0$ ($i=1,2,3,n$),
corresponding to the isotropic general relativistic vacuum solution
(Minkowski spacetime) which we are not interested in here.
The variable $\lambda$ runs from $-\infty$ to $+\infty$,
which means that $t - t_0$ is on the positive or negative half-line,
depending on the sign of $C_1 + C_2 + C_3$,
\ba
-\infty < t < t_0 &{\ \ \rm for\ \ }& C_1 + C_2 + C_3 < 0\l{tless}\\
t_0 < t  < \infty &{\ \ \rm for\ \ }& C_1 + C_2 + C_3 > 0. \l{tmore}
\ea
Henceforth we will consider solutions only for
$C_1+C_2+C_3>0$ without loss of generality. When $C_1+C_2+C_3<0$ we
obtain the time-reversed solutions for $t<t_0$. For simplicity we
shall set $t_0 = 0$ below; it can be reintroduced by substituting $t -
t_0$ for $t$ in the appropriate expressions.

\subsection{Dilaton-vacuum solutions}

For purposes of comparison, let us first give the solutions for
$L_3=0$, when the axion field $h$
and tensor potential $B_{\mu\nu}$
remain constant and so do not affect the dynamics. This corresponds to
the well-known dilaton-vacuum cosmology \cite{Gas+Ven 93}. We
have $d\phi/d\lambda=C_1+C_2\pm C_0$ and thus
\be
e^\phi = \exp \left( \pm C_0(\lzero) -C_1(\lone) - C_2(\ltwo) \right)
\; ,
\ee
Thus we have from Eq.~(\ref{a1sol}--\ref{a3sol})
\ba
a_1 & = & \exp\left({\half C_1(\lone) - \half C_2(\ltwo)
\pm\half C_0(\lzero)}\right) , \\ a_2 & = & \exp\left(\half C_2(\ltwo)
- \half C_1(\lone) 			\pm\half C_0(\lzero) \right) ,
\\ a_3 & = & \exp(C_3(\lthree)) ,\\
\ea
The constants $C_0, C_i$ are constrained by Eq.~(\ref{czip}), while
the constants $\lambda_0, \lambda_i$ are free. All of these constants,
of course, are fixed by the initial conditions on the cosmology.

Using Eq.~(\ref{time}) to re-write these in terms of the proper time
in the string frame we have simply power-law solutions
\ba
e^\phi &=& e^{\phi_*} \tts^{\mp p+q-1} , \label{vacpsol} \\ a_1 &=&
a_{1*} \tts^{-\half(\pm p+r)} , \label{vaconesol} \\ a_2 &=& a_{2*}
\tts^{-\half(\pm p-r)} , \label{vactwosol} \\ a_3 &=& a_{3*}\tts^q
\label{vacthsol} .
\ea
Here we have renamed various combinations of constants, in particular
we have
\ba
p = {C_0\over {C_1 + C_2 + C_3}},
& \qquad & q = {C_3\over {C_1 + C_2 + C_3}}, \nonumber \\
r = {C_2 - C_1 \over {C_1 + C_2 + C_3}}
\qquad & {\rm and} & \qquad s = {C_n \over {C_1 + C_2 + C_3}}.\l{pqr}
\ea
The various prefactors $e^{\phi_*}, a_{i*}\ (i=1,2,3)$, are
appropriate combinations of the constants that appear in
$\lambda$-time solutions. The characteristic time $t_*$ corresponds
to the value of $t$ when $\lambda=\lambda_0$,
\be
\tts^\frac{1}{C_1 + C_2 + C_3}
= \exp(\lzero).\l{tstar}
\ee
In terms of these new constants the constraint, Eq.~(\ref{czip}),
becomes
\be
\half p^2 + q^2 + \half r^2 + ns^2= 1 \ .
\ee
Solutions with $p<0$ correspond to solutions for $t$ (and thus $t_*$)
negative. Considering only solutions for $t>0$ (i.e.~$C_1+C_2+C_3>0$)
implies that $p\geq0$.  In either case we have two possible vacuum
branches corresponding to the choice of $\pm p$ in the solutions
(unless $C_0$, and thus $p$, are zero).

While these solutions are in general anisotropic, this is simply a
consequence of having allowed ourselves the freedom to choose
anisotropic initial conditions. If we pick isotropic initial
conditions the metric remains an isotropic
(Friedmann-Robertson-Walker) metric.

\subsection{Axion-dilaton solutions}

When $L_3\neq0$ we define $v = d\phi/d\lambda + (C_1 + C_2)$ so that
the equation of motion for the dilaton, Eq.~(\ref{diltwo}), can be
written as
\be
{dv\over d\lambda} = - e^{2\phi}L_3^2\ta_1^2\ta_2^2 = v^2 -
C_0^2,\l{veq}
\ee
which can be solved to give (note that $v^2 < C_0^2$)
\be
v = - C_0\tanh C_0(\lzero) \l{vsol} ,
\ee
where $\lambda_0$ is a constant of integration. The solution for the
dilaton then follows from Eq.~(\ref{veq}),
\be
e^\phi = \bigg|{C_0\over L_3}\bigg|{1\over \cosh C_0(\lzero)}\exp(-
C_1(\lone) - C_2(\ltwo)) \qquad {\rm for}\ L_3\neq0 .\l{dilsol}
\ee
We can now collect our solutions for the scale factors in the string
frame,
\ba
a_1 &=& \sqrt{\left|{C_0\over L_3}\right|} {\exp\big({\half C_1(\lone)
- \half C_2(\ltwo)}\big)
\over\sqrt{\cosh C_0(\lzero)}}
,\l{onesol}\\
a_2 &=& \sqrt{\left|{C_0\over L_3}\right|}
{\exp\big(\half C_2(\ltwo) - \half C_1(\lone)\big)
\over\sqrt{\cosh C_0(\lzero)}}
,\l{twosol}\\
a_3 &=& \exp(C_3(\lthree)).\l{threesol}
\ea
Note how the dilaton-vacuum solutions contain two distinct branches
according to whether we choose $\pm C_0$, whereas the axion-dilaton
results above are independent of the choice of sign, smoothly evolving
from the $-C_0$ vacuum branch, when $(\lzero)$ is large and negative,
to the $+C_0$ branch, when $(\lzero)$ becomes large and positive.

In terms of the string frame time coordinate $t$, the solutions for
the scale factors and the dilaton take the following forms, [with $p$,
$q$, and $r$ as defined in Eq.~(\ref{pqr})]
\ba
e^\phi &=& e^{\phi_*} \left[\tts^{p-q+1} + \tts^{-p-q+1}\right]^{-1}\;
,
\l{tpsol}\\
a_1 &=& a_{1*}\left[\tts^{p+r} + \tts^{-p+r}\right]^{-\half},
\l{tonesol}\\
a_2 &=& a_{2*}\left[\tts^{p-r} + \tts^{-p-r}\right]^{-\half},
\l{ttwosol}\\
a_3 &=& a_{3*}\tts^q.\l{tthsol}
\ea
Again we see the evolution from one vacuum branch (with the lower signs
in Eqs.~(\ref{vacpsol}--\ref{vacthsol})) for $t\ll t_0$ to the other
vacuum branch (with upper signs) for $t\gg t_*$.

Note that the effect of the axion field is to decelerate the scale
factors $a_1$ and $a_2$, placing an upper bound on the product
\be
a_1 a_2 \; = \; {a_{1*} a_{2*} \over \tts^p + \tts^{-p}} \;
 \leq \; \left| {C_0 \over L_3} \right| \; .
\ee
In contrast with the dilaton-vacuum solutions, even an initially
isotropic metric ($r=0$, $q=p/2$) becomes anisotropic in the presence
of the axion resulting from a time-dependent tensor potential. The
only solution that can approach isotropy at late times is a {\em
contracting} metric.

To understand how this occurs it is useful to return to the Einstein
frame solutions. We can always write the Einstein constraint,
Eq.~(\ref{ancon}), in a Bianchi type I metric as
\be
\ttheta^2 = 3 \kappa^2 \left( \trho_\phi+\trho_\beta+\trho_H \right)
 + 3\tsigma^2
 \label{shearcon}
\ee
where
\ba
\ttheta^2 & \equiv & \left( \dot\tal_1 + \dot\tal_2 + \dot\tal_3
 \right)^2 \\
 & = & {1 \over 4\ta_1^2\ta_2^2\ta_3^2}
  \left\{ 3(C_1+C_2) + 2C_3 + C_0\tanh C_0(\lambda-\lambda_0)
 \right\}^2
\ea
is the expansion rate and the anisotropy (or ``shear'') is given by
\ba
\tsigma^2 & \equiv & {1\over3} \left( \dot\tal_1^2 + \dot\tal_2^2
 + \dot\tal_3^2 - \dot\tal_1\dot\tal_2
 - \dot\tal_2\dot\tal_3 - \dot\tal_3\dot\tal_1 \right) \\
& = & {1\over4\ta_1^2\ta_2^2\ta_3^2}
 \left\{ C_1^2+C_2^2+2(C_1-C_2)^2 +
   \left[2C_3 + C_0\tanh C_0(\lambda-\lambda_0)\right]^2 \right\}
  \label{shear}
\ea
for the axion-dilaton solutions given in
Eqs.~(\ref{onesol}--\ref{threesol}).

Each term in the constraint Eq.~(\ref{shearcon}) is non-negative and
so the
relative importance of each term on the right-hand-side in determining
the expansion rate is simply given by their magnitude.
Thus at early or late times we recover the
dilaton-vacuum solutions where the expansion rate is proportional to
square of the volume in the Einstein frame
($\ttheta^2\propto\trho_\phi\propto\trho_\beta\propto\tsigma^2$
as $\lambda\to\pm\infty$),
while the axion energy density evolves as
\ba
\trho_H & = & {L_3^2\over4\kappa^2}
              {a_1^2a_2^2 \over \ta_1^2\ta_2^2\ta_3^2} \\
 & = & {C_0^2\over4\kappa^2\ta_1^2\ta_2^2\ta_3^2}
	\left\{ 1 - \tanh^2C_0(\lambda-\lambda_0) \right\}
\ea
and vanishes relative to the other terms as $\lambda\to\pm\infty$.

The axion field only plays a dynamical role for a
brief period around $\lambda\simeq\lambda_0$ ($t\simeq t_*$).
It is the only anisotropic fluid in the
system, so it delivers an ``anisotropic impulse'' to the metric.
As can be seen from Eq.~(\ref{shear}) this causes a change in the
shear, around $\lambda=\lambda_0$ ($t=t_*$), $\Delta\tsigma^2\propto
pq$. The only stable late time vacuum solutions have the area
perpendicular to the gradient of the axion field ($a_1a_2$) decreasing.

In order to solve for $B_{\mu\nu}$, we first note that we have already
chosen $B_{0i} = 0$, and setting $B_{ij} \equiv B_{ij}(t)$ implies
$H_{ijk} = 0$.  Further, the choice $L_1 = L_2 = 0$, made because of
Eq.~(\ref{LiLj}), implies that $B_{23}$ and $B_{31}$ are constants, and
it follows that $H_{012} = \del_t B_{12}$. Then, combining the two
expressions
\ba
\tH^{012} &=& -{e^\phi\over\ta_1^2\ta_2^2}\del_t B_{12} \l{tHone}\\
\tH^{012} &=& {e^{2\phi}L_3\over\sqrt{\det\tg}}\l{tHtwo}
\ea
and using the definition of $\lambda$, Eq.~(\ref{deflam}), we obtain
\be
\del_\lambda B_{12} = - L_3e^{2\phi}\ta_1^2\ta_2^2 = - L_3a_1^2a_2^2.
\l{delb}
\ee
Of course, if $L_3=0$, then $B_{12}$ is just a constant.  Otherwise,
use of Eqs.~(\ref{a1sol}), (\ref{a2sol}) and~(\ref{dilsol}) allows us
to rewrite the above equation and solve for $B_{12}$,
\be
\del_\lambda B_{12} = - {C_0^2\over L_3}{1\over \cosh^2 C_0(\lzero)}
\qquad \Rightarrow \qquad B_{12} = - {C_0\over L_3}\left(\tanh
C_0(\lzero) + \bzero\right),\l{Bsol}
\ee
where $\bzero$ is a constant fixed solely by the initial conditions on
the antisymmetric tensor and is independent of the choice of all the
other constants. For the sake of completeness, the other components of
$B_{\mu\nu}$ are given by $B_{0i} = 0, B_{23} = \widehat B_{23},
B_{31} = \widehat B_{31}$, both constants.

Thus we see that, except for
$\lambda\sim\lambda_0$ (or equivalently $t\sim t_*$), the tensor
potential remains very nearly constant and we recover the vacuum
solutions. Only in the vicinity of $\lambda=\lambda_0$ does the
potential change, resulting in a non-zero axion field, which delivers
an anisotropic impulse to the metric, before the potential becomes
roughly constant again returning to the vacuum branch.

\section{Duality}

The class of homogeneous solutions of the metric, dilaton and
antisymmetric tensor potential in four dimensions has been shown to
have a global $O(3,3)$ invariance (in general, a global $O(D-1, D-1)$
invariance in $D$ dimensions) \cite{Mei+Ven 91,Sen 91,duality} under
which
\be
M \to M' = \Omega^TM\Omega,\qquad\qquad \bar\phi \equiv \phi -
\ln\sqrt{\det G}
\to \bar\phi,\l{dual1}
\ee
where $\Omega$ is a $6\times6$ constant matrix satisfying
\be
\Omega^T\eta\Omega = \eta, \qquad\qquad
\eta = \left(
\begin{array}{cr}
 {\bf 0}& {\bf 1}\\ {\bf 1}&{\bf 0}\end{array}
\right),\l{omega}
\ee
({\bf 1} is the $3\times 3$ identity matrix) and
\be
M \equiv \left(
\begin{array}{cc}
G^{-1}& - G^{-1}B\\ BG^{-1}& G - BG^{-1}B
\end{array}
\right),\l{mdef}
\ee
where $G$ and $B$ are respectively $g_{ij}$ and $B_{ij}$ written as
$3\times3$ matrices. Any $6\times 6$ constant matrix $\Omega$ obeying
Eq.~(\ref{omega}) generates new solutions for the metric, antisymmetric
tensor and the dilaton, corresponding to $M'$, from the original set
of solutions.

In the case where $B_{ij}$ vanishes the special choice $\Omega = \eta$
is called the `scale factor duality' transformation
\cite{Mei+Ven 91,Sen 91} because it takes the scale factors
$a_i \to a_i^{-1}$, thus exchanging `large' directions with `small'
directions. Let us consider what happens to the solutions of the
previous section with this particular choice of $\Omega$.
This generates non-equivalent solutions, i.e., it is not just a
spatial rotation or gauge transformation \cite{Sen 91}, while it will
allow us remain within our class of Bianchi I solutions if we set
$\widehat B_{23} = \widehat B_{31} = 0$.

In this case, a bit of algebra shows (with primes denoting the duality
transformed functions)
\ba
{a'}_1^2 &=& {a_2^2\over a_1^2a_2^2 + B_{12}^2}\l{a1du1}\\ {a'}_2^2
&=& {a_1^2\over a_1^2a_2^2 + B_{12}^2}\l{a2du1}\\ {a'}_3^2 &=& {1\over
a_3^2}\l{a3du1}\\ B'_{12} &=& -{B_{12}\over a_1^2a_2^2 +
B_{12}^2}.\l{bdu1}
\ea
{}From our solutions (\ref{onesol}), (\ref{twosol} and (\ref{Bsol})
we have
\be
a_1^2a_2^2 + B_{12}^2 = {C_0^2\over L_3^2\cosh C_0(\lzero)}
\left[(1 + \bzero^2)\cosh C_0(\lzero) +
2\bzero\sinh C_0(\lzero)\right].
\l{denom1}
\ee
It follows that the dual solutions can be divided into two separate
classes.

\noindent {\it(a)} $\bzero \neq \pm 1$. In this case, we can define a
constant $\lambda_0''$ such that
\be
2\bzero = b\sinh C_0\lambda_0'',\l{defbl''}
\ee
where $b = |1 - \bzero^2|$. Then $1 + \bzero^2 = b\cosh
C_0\lambda_0'',$ and Eq.~(\ref{denom1}) becomes
\be
a_1^2a_2^2 + B_{12}^2 = {bC_0^2\over L_3^2}{\cosh
C_0(\lzerop)\over\cosh C_0(\lzero)},\l{denom2}
\ee
with $\lambda'_0 = \lambda_0 - \lambda''_0$. The dual transformed
solutions (\ref{a1du1} -- \ref{bdu1}) can be then rewritten as
\ba
a'_1 &=& \sqrt{\left|{L_3\over bC_0}\right|}
{\exp\big({\half C_2(\ltwo) - \half C_1(\lone)}\big)
\over\sqrt{\cosh C_0(\lzerop)}}
,\l{a1du2}\\
a'_2 &=& \sqrt{\left|{L_3\over bC_0}\right|}
{\exp\big(\half C_1(\lone) - \half C_2(\ltwo)\big)
\over\sqrt{\cosh C_0(\lzerop)}}
,\l{ta2du2}\\
a'_3 &=& \exp(- C_3(\lthree))\l{a3du2}\\
B'_{12} &=& \mp {L_3\over bC_0}\left(\tanh C_0(\lzerop)
- \bzero\right),\l{bdu2}
\ea
where the sign for $B'_{12}$ is positive or negative as $(1 -
\bzero^2)$ is negative or positive, respectively. As is obvious, these
fall in the same classes of solutions as our original ones
Eqs.~(\ref{dilsol} -- \ref{threesol}, \ref{Bsol}). The dilaton $\phi$
is shifted,
\be
e^\phi \to e^{\phi'} = \left|{L_3\over bC_0}\right|{1\over\cosh
C_0(\lzerop)}\exp[- C_1(\lone) - C_2(\ltwo) - 2C_3(\lthree)],\l{duphi}
\ee
while $\bar\phi$ as defined in Eq.~(\ref{dual1}) is invariant, as is
the coordinate $\lambda$. Obviously, the duality transformation can be
expressed as a transformation on the constants that appear in the
solutions.  For example, when the solutions are expressed in terms of
the string time coordinate $t$ as in Eqs.~(\ref{tpsol}--\ref{tthsol}),
the duality transformation is essentially equivalent to the following
transformation on the constants:
\be
p\to -p, \qquad q\to -q, \qquad r\to -r, \qquad
 L_3\to \pm |1-\bzero^2| {C^2_0\over L_3}\; .\l{ttrans}
\ee
Note that the duality transformation changes the
characteristic time defined by Eq.~(\ref{tstar}), (but now with
$\lambda_0$ replaced by $\lambda'_0$)
\be
t_* \to t'_* = t_*e^{-(C_1 + C_2 + C_3)\lambda_0''} = t_*\left|{{1 -
\bzero}\over {1 + \bzero}}\right|^{1/p}.
\l{tsttr}
\ee
Thus the characteristic time tends to zero or infinity as $\bzero$
tends to $+1$ or $-1$ respectively.

\noindent{\it(b)} $\bzero = \pm 1$. In this case, we have
\ba
a_1^2a_2^2 + B^2_{12} &=& {C_0^2\over L_3^2}\left(\sech^2C_0(\lzero) +
(\tanh C_0(\lzero) \pm 1)^2\right)\nonumber\\ &=& {C_0^2\over
L_3^2}{2e^{\pm C_0(\lzero)}\over\cosh C_0(\lzero)}.
\l{denom3}
\ea
The dual transformed solutions turn out to be nothing more than
anisotropic solutions of pure dilaton cosmology,
\ba
a'_1 &=& \sqrt{\left|{L_3\over 2C_0}\right|}
\exp\big(\half C_2(\ltwo) - \half C_1(\lone)\mp
\half C_0(\lzero)\big),\l{a1du3}\\
a'_2 &=& \sqrt{\left|{L_3\over 2C_0}\right|}
\exp\big(\half C_1(\lone) - \half C_2(\ltwo)\mp
\half C_0(\lzero)\big),\l{ta2du3}\\
a'_3 &=& \exp(- C_3(\lthree))\l{a3du3}\\
B'_{12} &=&  {L_3\over 2C_0}\left(\sinh C_0(\lzero) \pm
\cosh C_0(\lzero)\right)e^{\mp C_0(\lzero)} =
\pm {L_3\over 2C_0}.\l{bdu3}
\ea
The sign ($\pm$) in the above corresponds to the sign of $\bzero$.
Thus we see that these `vacuum' ($H_{\mu\nu\lambda}$ = 0) solutions
appear as a limit of the duality transforms of type $(a)$ above as
$\lambda''_0 \to\pm\infty$ (alternatively, as $t'_* \to 0$ or
$\infty$).

\section{Conclusions}

We have considered four-dimensional cosmological solutions of low
energy effective string theory in which the metric, dilaton, modulus
and antisymmetric tensor potential depend only on the background time
coordinate. This restriction inevitably leads us to a homogeneous but
anisotropic universe except for the isotropic vacuum solution with the
antisymmetric tensor $B_{\mu\nu} = $const. Even in a situation with
$B_{ij}\neq0$ (but with $H_{\mu\nu\lambda} = 0$) the three components
of $B_{ij}$ form a three-vector, thus specifying a chosen direction.
However it is when the components of $B_{\mu\nu}$ are allowed to vary
with time that the variation (the axion field $h$) drives the
anisotropy of the universe.  In order to gain an understanding of how
such a universe would evolve, we have considered Bianchi type I
universes in this article possessing shear but no spatial curvature.

We find models which behave like dilaton-vacuum models (where the
axion can be neglected) at early and late times.  However the axion
field does affect the dynamics for a brief period around $t=t_*$
producing an anisotropic ``impulse'' at this point.  The Einstein
metric provides a useful frame in which to discuss the behaviour of
these solutions. During the effectively vacuum regimes, the shear,
$\tsigma^2$, and density of the modulus, $\trho_\beta$, and dilaton,
$\trho_\phi$, in this frame drive the expansion and are proportional
to $(\ta_1\ta_2\ta_3)^{-2}$. However the axion density,
$\trho_H\propto e^{2\phi}\ta_3^{-2}$, will grow relative to the shear,
modulus and dilaton densities at early times while
$e^{2\phi}\ta_1^2\ta_2^2= a_1^2a_2^2$ grows.  Thus the axion's
anisotropic pressure must eventually become important. It tends to
decelerate the scale factors $a_1$ and $a_2$, and produces an upper
bound on the product $L_3a_1a_2 \leq C_0$. The dilaton-vacuum
solutions ($L_3=0$) are thus atypical of the general axion-dilaton
solutions. The stable, late time, effectively vacuum era must have
$a_1a_2$ decreasing, while the third scale factor $a_3$ is free to
grow (or decrease) monotonically.

Such anisotropy is not observed in our universe today. We usually
expect shear will be diluted away at late times by the presence of
other isotropic matter, in particular isotropic radiation in the hot
big bang model.  However it is far from clear whether ordinary
isotropic radiation can dominate over the axion in this model. The
isotropic late-time behaviour (with $a=a_1=a_2=a_3$) dominated by a
radiation ($p=\rho/3$) is only possible if $\trho_H/\trho\propto
e^\phi a^2\to0$ as $t\to0$.  Because the pure radiation plus dilaton
solution has $\phi\to$const, a late time isotropic radiation-dominated
solution must be a {\em contracting} universe, i.e.~$a\to0$.

We cannot rule out some expanding isotropic solution with radiation at
late times, but for the anisotropic axion field not to spoil this
isotropy we must have $\trho_H/\trho$ remaining negligible, and thus a
decreasing dilaton, or some reason (such as an inflationary era in the
Einstein frame) for the axion gradient to be vanishingly small.  As
any variation from the standard hot big bang model (with constant
dilaton) is tightly constrained by, for instance, results from
primordial nucleosynthesis, radiation alone does not seem to be
sufficient to erase the anisotropic influence of a time-dependent
antisymmetric tensor potential.  Spatial curvature (zero in the
Bianchi type I metric) would in general introduce further anisotropy.
Again, we would require inflation in the Einstein frame to avoid
curvature dominating the evolution at late times.

The `characteristic time' $t_*$ plays a major role in both the
evolution and in the interpretation of duality transformations of the
solutions.  At early times, the antisymmetric tensor $B_{\mu\nu}$ is
approximately constant. It changes rapidly around $t_*$ and becomes
approximately constant again, albeit at a different value.  When a
transformed characteristic time $t'_*$ is defined, the duality
transformation is seen to change the time dependence of the scale
factors and the antisymmetric tensor potential by
$\left(t/t_*\right)\to\left(t'_*/ t\right)$. The duality
transformation is then seen to relate a given solution at late times
with another solution at early times. In particular, we find that for
special choices of initial values of the antisymmetric tensor, the
universe that results is duality related to a vacuum solution of pure
dilaton cosmology, where $t'_*\to0$ or $\infty$.

The $O(3,3)$ invariance of the low-energy action proves to be of
limited use in a cosmological context. The requirement that the metric
and potential $B_{\mu\nu}$ both be functions only of time is highly
prescriptive. Given a homogeneous metric a more natural expectation
would be that the axion field (which determines the energy-momentum
tensor and thus the metric) should be time-dependent \cite{CLW/6/94},
rather than the potential. The only metric (with zero-vorticity) which
can meet this presciption is Bianchi type I and in such a case the
axion field derived from $B_{\mu\nu}(t)$ can have no time-dependence
and must be anisotropic. This prohibits isotropic expanding universes
at late times in the string frame. In this respect the dilaton-vacuum
solutions $B_{\mu\nu}=$const are atypical of the behaviour of the
general axion-dilaton solutions.

\section*{Acknowledgements}

The authors are supported by the PPARC.
EJC would like to thank Tom Kephart, Tanmay Vachaspati and Jnanadeva
Maharana for helpful discussions.


\nonfrenchspacing


\begin{thebibliography}{99}
\frenchspacing

\bibitem{Gas+Ven 93}
M. Gasperini \& G. Veneziano, {\sl Astroparticle Physics} {\bf 1} 317
(1993)

\bibitem{dilaton}
M. Mueller, {\sl Nucl. Phys.} {\bf B337} 37 (1990);
I. Antoniadis, C. Bachas, J. Ellis \& D. V. Nanopoulos, {\sl Phys.
 Lett.} {\bf B257} 278 (1991);
A. A. Tseytlin, {\sl Int. J. Mod. Phys. D} {\bf 1} 223 (1992);
A. A. Tseytlin \& C. Vafa, {\sl Nucl. Phys.} {\bf B372} 443 (1992);
N. Kaloper and K. Olive, {\sl Astroparticle Physics} {\bf 1} 185 (1993);
R. Brustein \& G. Veneziano, {\sl Phys. Lett.} {\bf B329} 429 (1994)

\bibitem{CLW/6/94}
E. J. Copeland, A. Lahiri \& D. Wands, {\sl Low energy effective string
 cosmology} preprint hep-th/9406216, {\sl to appear Phys. Rev. D}

\bibitem{axion}
D. S. Goldwirth and M. J. Perry, {\sl Phys. Rev. D}{\bf 49} 5019 (1994);
K. Behrndt and S. F\"orste, hep-th/9403179, {\sl to appear Nucl. Phys.}

\bibitem{SL2R}
A. Shapere, S. Trivedi and F. Wilczek, {\sl Mod. Phys. Lett.} {\bf A6}
2677 (1991); A. Sen, {\sl ibid} {\bf A8} 2023 (1993)

\bibitem{Mei+Ven 91}
K. A. Meissner \& G. Veneziano, {\sl Mod. Phys. Lett.} {\bf A6} 3397
(1991)

\bibitem{Sen 91}
A. Sen, {\sl Phys. Lett.} {\bf B271} 295 (1991)

\bibitem{effaction}
E. S. Fradkin \& A. A. Tseytlin, {\sl Nucl. Phys.} {\bf B261} 1
(1985);
C. G. Callan, D. Friedan, E. J. Martinec \& M. J. Perry, {\sl
Nucl. Phys.} {\bf B262} 593 (1985);
C. Lovelace, {\sl Nucl. Phys.} {\bf B273} 413 (1985)

\bibitem{MTW}
C. W. Misner, K. S. Thorne \& J. A. Wheeler, ``Gravitation'', W. H.
Freeman \& Company, New York (1973)

\bibitem{duality}
K. S. Narain, {\sl Phys. Lett.} {\bf B169} 41 (1986); K. S. Narain, M.
H. Sarmadi \& E. Witten, {\sl Nucl. Phys.} {\bf B279} 369 (1987); A.
Shapere \& F. Wilczek, {\sl Nucl.  Phys.} {\bf B320} 669 (1989); K. A.
Meissner \& G. Veneziano, {\sl Phys. Lett.} {\bf 267} 33 (1991).

\end{thebibliography}
\end{document}